# AI Meets Maritime Training: Precision Analytics for Enhanced Safety and Performance

Vishakha Lall [1*], Yisi Liu [1]

[1] Centre of Excellence in Maritime Safety, Singapore Polytechnic, Singapore
* Corresponding author: e-mail: vishakha_lall@sp.edu.sg

Traditional simulator-based training for maritime professionals is critical for ensuring safety at sea but often depends on subjective trainer assessments of technical skills, behavioral focus, communication, and body language, posing challenges such as subjectivity, difficulty in measuring key features, and cognitive limitations. Addressing these issues, this study develops an AI-driven framework to enhance maritime training by objectively assessing trainee performance through visual focus tracking, speech recognition, and stress detection, improving readiness for high-risk scenarios. The system integrates AI techniques, including visual focus determination using eye tracking, pupil dilation analysis, and computer vision; communication analysis through a maritime-specific speech-to-text model and natural language processing; communication correctness using large language models; and mental stress detection via vocal pitch. Models were evaluated on data from simulated maritime scenarios with seafarers exposed to controlled high-stress events. The AI algorithms achieved high accuracy, with ~92% for visual detection, ~91% for maritime speech recognition, and ~90% for stress detection, surpassing existing benchmarks. The system provides insights into visual attention, adherence to communication checklists, and stress levels under demanding conditions. This study demonstrates how AI can transform maritime training by delivering objective performance analytics, enabling personalized feedback, and improving preparedness for real-world operational challenges.

## 1. INTRODUCTION

Traditional simulator-based training has long been a cornerstone in preparing maritime professionals, offering a controlled environment to practice and refine essential skills. This method allows trainees to engage in realistic scenarios, enhancing their technical proficiency and decision-making abilities without the immediate risks associated with real-world operations. However, despite its widespread adoption, this training approach is not without limitations.

A key challenge in performance assessments is the subjectivity of instructor evaluations, which rely on personal judgment to assess technical skills, communication, and behaviour. This approach can lead to inconsistencies and biases, as interpretations vary among instructors. Additionally, the simultaneous monitoring of multiple competencies can overwhelm instructors, increasing the risk of oversight and reducing assessment accuracy. (*Malik A.A. & Zafar Nargus, 2015*)

Traditional simulator-based training lacks tools to objectively measure critical factors like visual focus and stress levels, which are vital in high-risk maritime scenarios where situational awareness and emotional regulation impact safety. Without precise measurement, assessments may miss key indicators of trainee readiness. While simulation training provides immersive experiences, studies highlight its limitations in capturing complex human factors, such as stress responses and attention management, essential for maritime operations. (*Emad Gholam Reza & Kataria Aditi, 2022*)

Addressing these challenges requires an objective, scalable, and scientifically grounded approach to improve maritime safety and training. AI integration offers a promising solution by enabling real-time, objective assessments of trainee performance.

## 2. RELATED WORK

### 2.1.1. Artificial Intelligence in Simulator Based Training

The aviation sector has extensively explored AI-enhanced training methods. For instance, (*Guevarra Michael & Das Srijita & Wayllace Christabel et al., 2023*) presented an intelligent tutoring framework that autonomously trains pilots in flight simulators using a simulated teacher. This system provides visual feedback to correct errors, reducing reliance on human instructors and enhancing learning efficiency. Similarly, (*Yang Shuiqiao & Yu Kun & Lammers Thorsten et al., 2021*) developed a real-time feedback framework for pilots, utilizing supervised models trained on standard operational patterns for various manoeuvres. These systems demonstrate AI's ability to offer tailored, data-driven feedback, significantly improving pilot competency.

In another groundbreaking study, (*Mohan Dilli Babu & Divya Venkatesh Jeevithashree & Prabhakar Gowdham et al., 2019*) estimated pilots' cognitive load in real-time using ocular parameters such as fixation, saccades, and pupil dilation. This approach provides deeper insights into pilots' mental workload, helping identify stress points and adapt training scenarios accordingly. These advancements in aviation highlight AI's capacity to augment traditional training by addressing cognitive and operational aspects.

### 2.1.2. Artificial Intelligence in Maritime Training

In maritime training, researchers are leveraging AI to enhance simulator-based learning. For instance, (*Munim Z & Kim T, 2023*) applied anomaly detection techniques to simulation log data, identifying performance deviations.

Eye-tracking technology has emerged as another valuable tool. (*Atik Oguz & Arslan Ömer, 2019*) employed heatmaps and comparative analysis between novice and expert ship officers to study differences in visual focus, aiding skill development in electronic navigation.

AI-driven speech analysis is also advancing maritime training. (*Jatta Lamin, 2024*) integrated text embeddings and large language models to improve Automated Speech Recognition (ASR) accuracy in maritime communication, ensuring precise interactions during high-stress scenarios.

Cognitive load and stress analysis represent critical applications of AI in this domain. (*Žagar Dejan et al., 2020*) explored cognitive load during simulated navigational tasks, using metrics like pupil diameter, electrodermal activity, and heart rate. Similarly, (Xue Helene et al., 2024) linked stress levels, measured through bio-signals, with training outcomes, finding that stress induced by reduced visibility or equipment failures impairs performance.

## 3. METHODOLOGY

## 3.1. Dataset

In our study, we utilize two datasets: one for training, fine-tuning and validating the various models, and another for evaluating our framework to demonstrate its benefits.

### 3.1.1. Training and Validation Dataset

For the vision dataset, participants wore Tobii Pro Glasses 3, which recorded egocentric video and gaze data as they navigated the simulator, intentionally focusing on various panels and equipment while varying their distance from these objects. This data was manually annotated to include panel and sub-panel detection and segmentation. For the audio dataset, participants with diverse South-East Asian accents read a conversational-style script related to maritime activities and were recorded using the integrated microphone of Tobii Pro Glasses 3. This dataset was manually annotated using timestamps and the script as transcriptions. For the stress dataset, the DAIC-WOZ (*Gratch Jonathan & Arstein Ron & Lucas Gale et al., 2014*) open-source dataset serves as a benchmark for validating models on distress.

### 3.1.2. Evaluation Dataset

The data was collected at Advanced Navigation Research Simulator at *Centre of Excellence in Maritime Safety, Singapore*. The participant performed two simulation-based navigation exercises one with good visibility and the other with reduced visibility conditions. During these exercises, demanding events such as potential collisions, engine failures, and squalls were introduced, simulating high-stress scenarios to assess participants' responses in dynamic maritime environments. The participant wore Tobii Pro Glasses 3, which recorded egocentric video and gaze data, along with an integrated microphone for audio capture.

## 3.2. Visual Focus

### 3.2.1. Eye Tracking

Eye-tracking technology monitors eye positions and movements to pinpoint where a subject is focusing. Modern wearable eye trackers, resembling regular eyeglasses, enable free movement while capturing gaze data. These devices integrate scene cameras for recording the user's perspective, infrared illuminators for precise gaze detection, and cameras to track pupil and corneal reflections. Advanced image processing triangulates these reflections, estimating the gaze vector and direction in 3D space. The gaze data is synchronized with the scene video, enabling real-time analysis of eye movements. Key features relevant to this study include:

- *Gaze Position* determines the horizontal and vertical coordinates of the gaze on the scene frame.
- *Gaze Depth* provides information about the distance of the object in focus.
- *Pupil Diameter* records the diameter of the pupil, which is an indicator of changes in cognitive attention. (*Reimer Jacob & Froudarakis Emmanouil & Cadwell Cathryn et al., 2014*)
- *Gaze Direction* represents the trajectory of eye movement and indicates where the user is looking relative to their head position.

### 3.2.2. Panel and Subpanel Detection Model

Using the egocentric perspective from eye-tracker glasses and corresponding gaze data, we developed an object detection model to identify the panel or equipment a subject focuses on in each frame. This system fine-tunes a pre-trained Vision Transformer (ViT) model (*Dosovitskiy Alexey & Beyer Lucas & Kolesnikov Alexander et al., 2020*), originally trained on ImageNet-21k with 14 million images and ~21,000 classes, and fine-tuned on the training dataset.

Panels and equipment, such as the Ship Management System or ECDIS, often feature large screens with multiple layers of information. To refine detection, the bounding box identified by the panel detection algorithm is used as a cropped frame for subpanel identification.

For this task, we employ a SegViT model (Zhang Bowen & Tian Zhi & Tang Quan et al., 2022), pre-trained on the ADE20k dataset for diverse semantic segmentation tasks. The model is fine-tuned on the training dataset. This approach enables precise subpanel detection within the bounding box, allowing for granular analysis of specific areas of interest in the subject's focus.

### 3.2.3. Attentional Focus

Research by (*Reimer Jacob & Froudarakis Emmanouil & Cadwell Cathryn et al., 2014*) suggests that pupil dilation is a reliable indicator of cognitive attention. Building on this, we define an attentional focus (AF) metric to differentiate between inattentive gazing and genuine focus on the gaze location by combining pupil dilation and gaze stability.

Pupil dilation ($PD$) reflects cognitive load and attentional state. The normalized pupil dilation, is calculated as:

$$PD_{norm} = \frac{PD - PD_{min}}{PD_{max} - PD_{min}} \quad (1)$$

Gaze stability ($GS$) measures how consistent the gaze point is across consecutive frames. Volatility in gaze is indicative of distractions or scanning behavior, while stable gazing suggests focused attention. It is computed using the Euclidean distance between gaze points across $N$ frames using gaze point $G$ and diagonal dimension of the screen $d_{screen}$:

$$GS = 1 - \frac{1}{N} \sum_{i=1}^{N-1} \frac{G_i - Gi + 1}{d_{screen}} \quad (2)$$

The attentional focus ($AF$) metric combines these two features using a weighted sum:

$$AF = w_1 PD_{norm} + w_2 GS \quad (3)$$

We validated our methodology and determined the weights $w_1$ and $w_2$ empirically on Ego Motion dataset. (*K. Ogaki & K. M. Kitani & Y. Sugano et al., 2012*)

By applying this methodology, we ensure that our attentional focus detection accurately reflects whether the subject is genuinely focusing or merely glancing at the gaze location.

### 3.3. Communication Analysis

### 3.3.1. Speech Recognition

OpenAI's Whisper model (*Radford Alec & Kim Jong & Xu Tao et al., 2022*) was designed as a versatile, plug-and-play speech recognition system, minimizing the need for extensive fine-tuning for specific applications.

While this broad applicability is beneficial in many contexts, it leads to reduced performance in specialized domains with constrained vocabularies, such as maritime communication. Governed by the Standard Maritime Communication Phrases (SMCP), maritime communication often faces inaccuracies when transcribed with the generic Whisper model, particularly with terms related to ports and locations.

To address this, we enhance Whisper's performance through contextual biasing, which adapts the model to prioritize maritime-specific vocabulary and reduce errors. Our model is fine-tuned on a maritime-specific vocabulary dataset (*Lall Vishakha & Liu Yisi, 2024*), constructed from audio recordings collected at the *Centre of Excellence in Maritime Safety, Singapore*, featuring speakers with various South-East Asian accents.

### 3.3.2. Communication Entity Extraction

Communication entity extraction identifies the recipient of a communication, revealing who the subject is alerting or informing. This is achieved through a fine-tuned Named Entity Recognition (NER) model.

We use a fine-tuned BERT (*Devlin Jacob & Chang Ming-Wei & Lee Kenton et al., 2018*) model for NER to detect and classify entities as internal (within the ship) or external (outside the ship). The model was fine-tuned on maritime communication datasets to recognize domain-specific entities like vessel names, port authorities, and onboard roles (e.g., captain, officer).

### 3.3.3. Communication Correctness

For triggered events like engine failures, predefined checklists outline required actions, such as contacting the engine room or preparing anchoring stations. These actions gauge the subject's readiness to handle emergency scenarios.

To assess checklist completion, we use a self-hosted pre-trained LLaMA 7B (*Touvron Hugo & Martin Louis & Stone Kevin et al., 2023*) Large Language Model (LLM). The checklists, created by trainers, serve as prompts for the model, which processes the subject's verbal responses and compares them to the checklist items. For each item, the model outputs a true/false value indicating whether the action was acknowledged or completed.

This method enables scalable, objective evaluation, leveraging the LLaMA model's contextual understanding to handle phrasing variations while aligning with trainer expectations.

## 3.4. Stress Detection

Traditional stress monitoring relies on bio-signals like heart rate variability and electrodermal activity, measured by wearable sensors. While accurate, these methods can be intrusive and uncomfortable. For simulation trainees, raw audio input offers an accessible, non-invasive alternative for stress detection, enabling real-time monitoring without the need for sensors.

Human communication involves non-verbal cues, such as rhythm, timbre, tone, and pitch, which are influenced by stress-related changes in muscle tension and respiration rate. (*Iu Hong & Frauendorfer Denise & Rabbi Mashfiqui et al., 2012*)

To capture these subtle variations, we use a transformer-based model to detect stress-induced acoustic changes in maritime training contexts.

## 4. RESULTS

## 4.1. Visual Focus

The results of the trained panel and subpanel detection model are presented in Table 1. For panels with multiple subpanels, the subpanel detection algorithm provides more granular insights, such as identifying 'Lateral Speed' within the 'Ship Management System.'

| Eye tracker frame with gaze (gaze fixation is plotted as a concentrated heatmap, arrows indicate gaze direction of the left and right eye) | Model Detection | Detection |
|---|---|---|
| 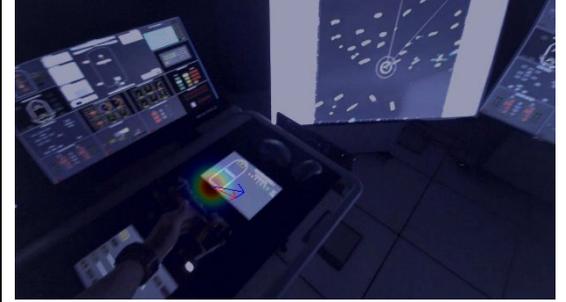 | 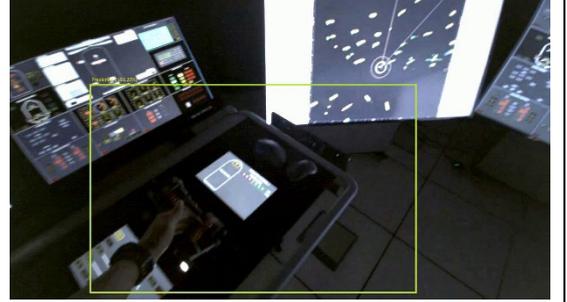 | Panel: Bow Thruster<br><br>Subpanel: NA |
| 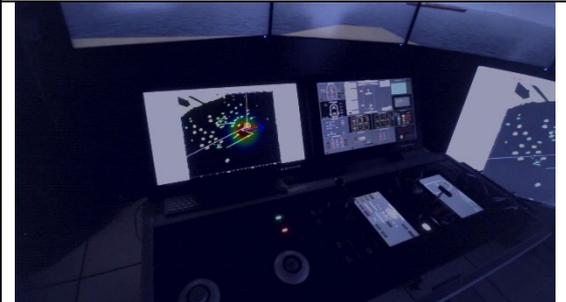 | 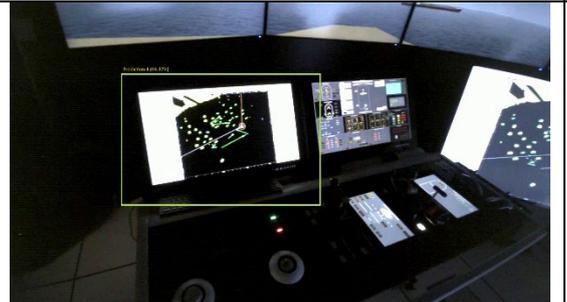 | Panel: Radar<br><br>Subpanel: Radar |
| 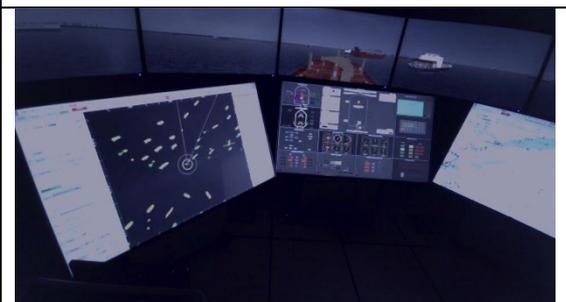 | 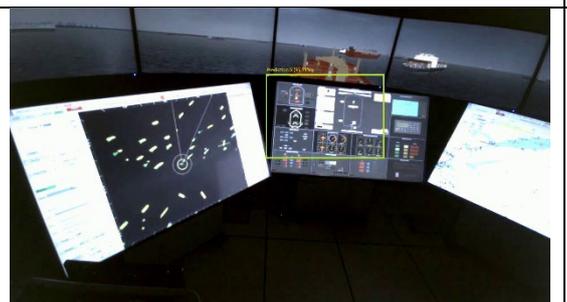 | Panel: Ship Management System<br><br>Subpanel: Lateral Speed |

*Table 1: Panel and Subpanel Detection Results*

Our method achieves panel detection performance of **95.04%** on the validation dataset.

## 4.2. Communication Correctness

Our approach significantly reduced the Word Error Rate (WER), demonstrating the impact of contextual adaptation. (*Lall Vishakha & Liu Yisi, 2024*)

Table 2 presents sample transcriptions on the test dataset produced by original Whisper model and contextually biased Whisper model against the ground truth. Upon closer examination, the qualitative benefits of contextual biasing become apparent. While the non-biased model exhibits erroneous transcriptions for locations such as `Keppel' and 'Brani', the biased model accurately identifies these locations.

(*Lall Vishakha & Liu Yisi, 2024*) demonstrates a classification accuracy of 98% on the validation dataset.

| Ground Truth from validation dataset | Transcriptions generated from original Whisper model | Transcription generated from contextually biased Whisper model |
|---|---|---|
| Keppel Control Keppel Control this is SMA Voyager, we are headed for Brani 7 and we have a vessel crossing ahead of us. Can you give us the name of that vessel over? | **Capital control, capital control**, this is SMA Voyager. We are headed for Brani's 7 and we have a vessel crossing a **herbivast**. Can you give us the name of that vessel over? | **Keppel Control, Keppel Control**, this Is Sma Voyager. we are headed for Brani 7 and We have a vessel crossing **ahead of us**. Can you give us the name of that vessel over? |
| Can you advise which berth is the vessel on my starboard side going to? Is it also berthing at brani over? | can you advise which **birth** is the vessel on my **star but side** going to is it also **birding** at **brownie** over? | Can You Advise which **berth** Is the Vessel On My **Starboard Side** Going to? Is It Also **berthing** At Brani Over? |

*Table 2: Sample transcriptions from original and improved Whisper model (Lall Vishakha & Liu Yisi, 2024)*

### 4.3. Stress Detection

Table 3 demonstrates that our model outperforms contemporary models, highlighting its superior capability in detecting stress on the validation dataset.

| Model | Accuracy |
|---|---|
| RNN | 76% |
| CNN+LSTM | 76% |
| CNN+GSOM | 89% |
| **Our model** | **90.33%** |

*Table 3: Performance metrics against contemporary models*

### 4.4. Framework Dashboard

The dashboard consolidates analysis across multiple modalities for each subject, providing a comprehensive evaluation of compliance with procedures, adherence to safety measures, and task correctness within the simulator environment. It enables trainers to pinpoint potential areas of weakness and compare performance across participants.

In the following results, we examine a case study involving a Main Engine Failure scenario for the subjects from the evaluation dataset. The subject was exposed to the same type of demanding event during two exercises: Exercise 1 with good visibility and exercise 2 under poor weather conditions with reduced visibility.

Official (Open)Official (Open)

Figure 1 depicts the subject's visual focus during the Main Engine Failure scenario, emphasizing behavioral differences between good and poor visibility conditions. Under low-visibility conditions, the subject notably reduces reliance on visual observation and increasingly depends on the ECDIS for navigation. Additionally, there is a heightened emphasis on the main engine screen, which is expected and critical during such a demanding event. These charts effectively assist trainers to identify abnormal visual focus patterns based on the navigation task and highlight instances where the subject may not allocate sufficient attention to critical equipment.

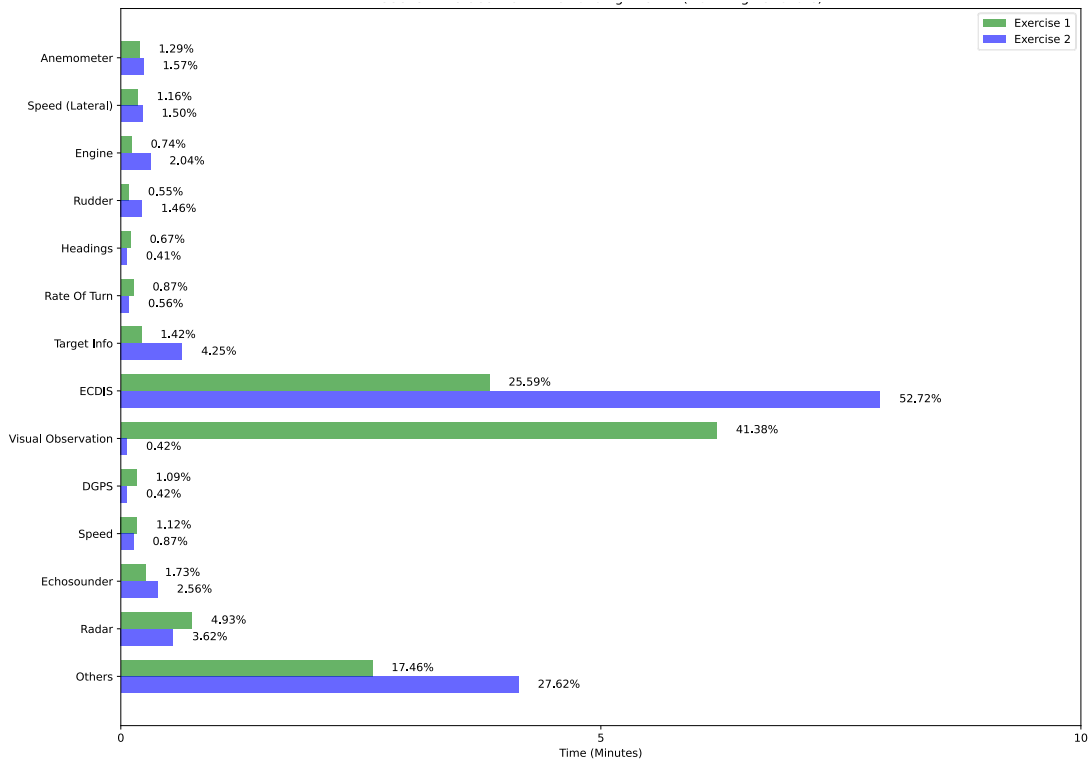

*Figure 1: Visual focus on various equipment in exercise 1 good visibility and exercise 2 poor visibility*

Figure 2 illustrates the subject's shift in attentional focus when the Main Engine Failure is triggered by the trainer. The focus intensifies sharply at the onset of the event and gradually diminishes as the situation stabilizes. These charts are instrumental in assessing whether the subject remains focused and attentive during training, providing insights into their potential performance during real-life scenarios.

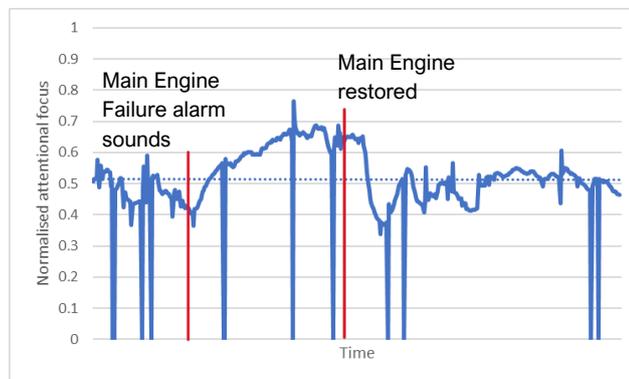

*Figure 2: Attentional focus of subject during Main Engine Failure*

Figure 3 illustrates the communication entities the subject interacts with, categorized as internal or external to the ship. In response to the engine failure, the subject communicates with the Engine Room and the Engineer to assess and address the issue. Under poor visibility conditions, there is an increased level of communication with the external entity, Port Control, reflecting a prudent and safety-oriented approach. These charts provide valuable insights for trainers, enabling them to review the exercise, ensure all necessary entities are engaged, and identify any safety-critical entities that may have been overlooked by the subject.

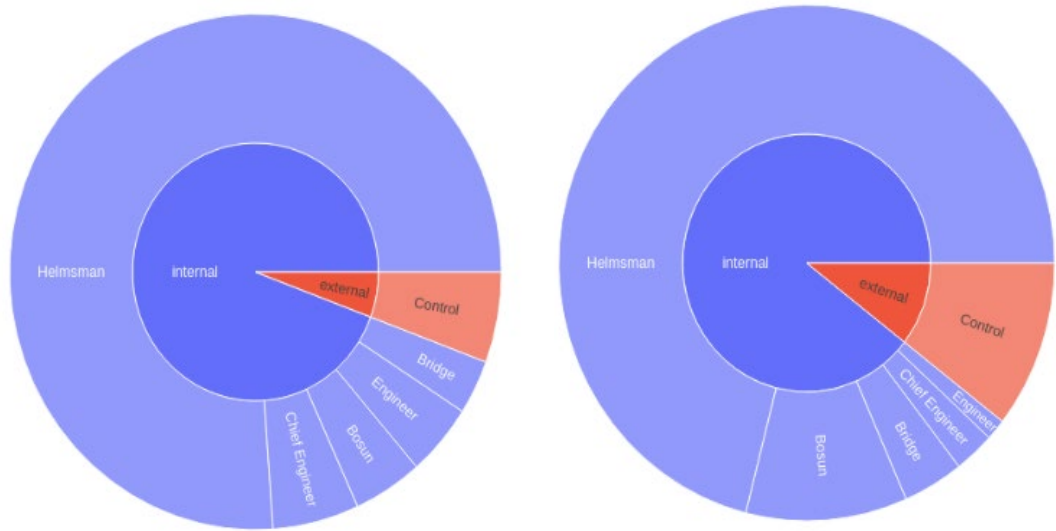

*Figure 3: Communication entities in exercise 1 good visibility (left) and exercise 2 poor visibility (right)*

Figure 4 provides deeper insights into the subject's communication patterns, revealing partial adherence to the required checklist guidelines for handling such scenarios. Specifically, the subject did not communicate with nearby ships to update them of an engine failure situation, nor kept the anchoring stations on standby which should be followed in a real scenario in cases when the engine does not get back up in time. Also, the checklist adherence builds upon the communication entity detection demonstrated earlier and even though the subject communicates with Port Control as highlighted Figure 3, they did not update them of the engine failure which is captured in the checklist adherence. Such charts highlight areas for improvement that trainers can objectively measure and provide as feedback to the subject.

| checklist | is_completed |
|---|---|
| Contacted engine room to know status | yes |
| Updated nearby vessels | no |
| Kept anchoring stations on stand by | no |
| Updated port control | yes |
| Contacted tug assistance | no |
| Contacted port marine safety | no |

*Figure 4: Checklist adherence during Main Engine Failure event*

Figure 5 illustrates the variation in the subject's stress levels throughout the exercise with 0 indicating no stress and 1 indicating stress, providing insights into their physiological response to the demanding event. A sharp increase in stress levels is observed immediately after the engine failure alarm is triggered. This heightened stress gradually subsides as the situation is brought under control and the subject resumes navigation. Stress levels recorded throughout the exercise offer trainers valuable insights into moments when the subject experienced anxiety, potentially triggered by the simulated danger or unfamiliarity of the scenario. These responses serve as an indicator of the subject's preparedness to handle similar situations in real life.

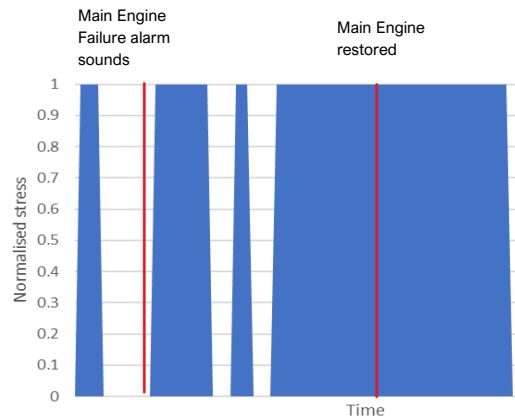

*Figure 5: Stress experienced by subject during Main Engine Failure*

## 5. CONCLUSION

This study demonstrates how AI can significantly enhance simulated maritime training by providing objective insights into trainee behaviour. Through personalized feedback and identification of key improvement areas, AI-driven analysis helps optimize training effectiveness and better prepares trainees for real-world operational challenges and potential mishaps.

## 6. ACKNOWLEDGEMENTS

We would like to express our sincere gratitude to the Singapore Maritime Institute (SMI) for their invaluable support and collaboration in providing the resources and expertise necessary for this research.

## REFERENCES

Atik, Oguz & Arslan, Ömer. (2019). Use of Eye Tracking for Assessment of Electronic Navigation Competency in Maritime Training. Journal of Eye Movement Research. 12. 10.16910/jemr.12.3.2.

Devlin, Jacob & Chang, Ming-Wei & Lee, Kenton & Toutanova, Kristina. (2018). BERT: Pre-training of Deep Bidirectional Transformers for Language Understanding. 10.48550/arXiv.1810.04805.

Dosovitskiy, Alexey & Beyer, Lucas & Kolesnikov, Alexander & Weissenborn, Dirk & Zhai, Xiaohua & Unterthiner, Thomas & Dehghani, Mostafa & Minderer, Matthias & Heigold, Georg & Gelly, Sylvain & Uszkoreit, Jakob & Houlsby, Neil. (2020). An Image is Worth 16x16 Words: Transformers for Image Recognition at Scale. 10.48550/arXiv.2010.11929.

Emad, Gholam Reza & Kataria, Aditi. (2022). Challenges of simulation training for future engineering seafarers - A qualitative case study. 10.54941/ahfe1002501.

Gratch, Jonathan & Arstein, Ron & Lucas, Gale & Stratou, Giota & Scherer, Stefan & Nazarian, Angela & Wood, Rachel & Boberg, Jill & DeVault, David & Marsella, Stacy & Traum, David & Rizzo, Albert & Morency, L.. (2014). The Distress Analysis Interview Corpus of human and computer interviews. (pp. 3123-3128).

Guevarra, Michael & Das, Srijita & Wayllace, Christabel & Demmans Epp, Carrie & Taylor, Matthew & Tay, Alan. (2023). Augmenting Flight Training with AI to Efficiently Train Pilots. Proceedings of the AAAI Conference on Artificial Intelligence. 37. 16437-16439. 10.1609/aaai.v37i13.27071.

Jatta, Lamin (2024). Maritime Automatic Speech Recognition : Improving the Quality of Transcriptions using Artificial Intelligence. https://urn.fi/URN:NBN:fi-fe2024060343553

K. Ogaki, K. M. Kitani, Y. Sugano and Y. Sato, "Coupling eye-motion and ego-motion features for first-person activity recognition," 2012 IEEE Computer Society Conference on Computer Vision and Pattern Recognition Workshops, Providence, RI, USA, 2012, pp. 1-7, doi: 10.1109/CVPRW.2012.6239188.

Lall, Vishakha & Liu, Yisi. (2024). Contextual Biasing to Improve Domain-specific Custom Vocabulary Audio Transcription without Explicit Fine-Tuning of Whisper Model. 1-6. 10.1109/MLNLP63328.2024.10800265.

lu, Hong & Frauendorfer, Denise & Rabbi, Mashfiqui & Mast, Marianne & Chittaranjan, Gokul & Campbell, Andrew & Gatica-Perez, Daniel & Choudhury, Tanzeem. (2012). StressSense: Detecting stress in unconstrained acoustic environments using smartphones. UbiComp'12 - Proceedings of the 2012 ACM Conference on Ubiquitous Computing. 351-360. 10.1145/2370216.2370270.

Mohan, Dilli Babu & Divya Venkatesh, Jeevithashree & Prabhakar, Gowdham & Saluja, Kamalpreet & Pashilkar, Abhay & Biswas, Pradipta. (2019). Estimating Pilots' Cognitive Load From Ocular Parameters Through Simulation and In-Flight Studies. Journal of Eye Movement Research. 12(3). 10.16910/jemr.12.3.3.

Munim, Z., Kim, T. (2023). A Review of Learning Analytics Dashboard and a Novel Application in Maritime Simulator Training. In: Salman Nazir (eds) Training, Education, and Learning Sciences. AHFE (2023) International Conference. AHFE Open Access, vol 109. AHFE International, USA. http://doi.org/10.54941/ahfe1003158

Radford, Alec & Kim, Jong & Xu, Tao & Brockman, Greg & McLeavey, Christine & Sutskever, Ilya. (2022). Robust Speech Recognition via Large-Scale Weak Supervision. 10.48550/arXiv.2212.04356.

Reimer, Jacob & Froudarakis, Emmanouil & Cadwell, Cathryn & Yatsenko, Dimitri & Denfield, George & Tolias, Andreas. (2014). Pupil Fluctuations Track Fast Switching of Cortical States during Quiet Wakefulness. Neuron. 84. 355-62. 10.1016/j.neuron.2014.09.033.

Touvron, Hugo & Martin, Louis & Stone, Kevin & Albert, Peter & Almahairi, Amjad & Babaei, Yasmine & Bashlykov, Nikolay & Batra, Soumya & Bhargava, Prajjwal & Bhosale, Shruti & Bikel, Dan & Blecher, Lukas & Ferrer, Cristian & Chen, Moya & Cucurull, Guillem & Esiobu, David & Fernandes, Jude & Fu, Jeremy & Fu, Wenyin & Scialom, Thomas. (2023). Llama 2: Open Foundation and Fine-Tuned Chat Models. 10.48550/arXiv.2307.09288.

Xue, Helene & Haugseggen, Øyvind & Røds, Johan-Fredrik & Batalden, Bjørn-Morten & Prasad, Dilip. (2024). Assessment of stress levels based on biosignal during the simulator-based maritime navigation training and its impact on sailing route reliability. Transportation Research Interdisciplinary Perspectives. 24. 101047. 10.1016/j.trip.2024.101047.

Yang, Shuiqiao & Yu, Kun & Lammers, Thorsten & Chen, Fang. (2021). Artificial Intelligence in Pilot Training and Education -Towards a Machine Learning Aided Instructor Assistant for Flight Simulators. Communications in Computer and Information Science. 10.1007/978-3-030-78642-7_78.

Žagar, Dejan & Svetina, Matija & Kosir, Andrej & Dimc, Franc. (2020). Human Factor in Navigation: Overview of Cognitive Load Measurement during Simulated Navigational Tasks. Journal of Marine Science and Engineering. 8. 10.3390/jmse8100775.

Zhang, Bowen & Tian, Zhi & Tang, Quan & Chu, Xiangxiang & Wei, Xiaolin & Shen, Chunhua & Liu, Yifan. (2022). SegViT: Semantic Segmentation with Plain Vision Transformers. 10.48550/arXiv.2210.05844.